\documentclass[aps,prl,twocolumn,superscriptaddress]{revtex4}

\usepackage{graphicx}

\begin{document}


\title{A model for the distribution of aftershock waiting times.}

\author{Robert Shcherbakov}
\email[]{roshch@cse.ucdavis.edu}
\affiliation{Center for Computational Science and Engineering,
University of California, Davis, CA 95616}

\author{Gleb Yakovlev}
\email[]{gleb@cse.ucdavis.edu}
\affiliation{Center for Computational Science and Engineering,
University of California, Davis, CA 95616}

\author{Donald L. Turcotte}
\email[]{turcotte@geology.ucdavis.edu}
\affiliation{Department of Geology, University of California, Davis,
CA 95616}

\author{John B. Rundle}
\email[]{rundle@cse.ucdavis.edu}
\affiliation{Center for Computational Science and Engineering,
University of California, Davis, CA 95616}

\date{July 28, 2005}

\begin{abstract}
In this work the distribution of inter-occurrence times between
earthquakes in aftershock sequences is analyzed and a model based on
a non-homogeneous Poisson (NHP) process is proposed to quantify the
observed scaling. In this model the generalized Omori's law for the
decay of aftershocks is used as a time-dependent rate in the NHP
process. The analytically derived distribution of inter-occurrence
times is applied to several major aftershock sequences in California
to confirm the validity of the proposed hypothesis.
\end{abstract}



\maketitle


The occurrence of earthquakes is an outcome of complex nonlinear
threshold dynamics in the brittle part of the Earth's crust. This
dynamics is a combined effect of different temporal and spatial
processes taking place in a highly heterogeneous media over a wide
range of temporal and spatial scales \cite{Rundle:2003a}. Despite
this complexity, one can consider earthquakes as a point process in
space and time, by neglecting the spatial scale of earthquake
rupture zones and the temporal scale of the duration of each
earthquake \cite{Vere-Jones:1970a,Ogata:1999a,Daley:2002a}. Then one
can study the statistical properties of this process and test models
that may explain the observed seismic activity.

In this letter, we analyze one of the important aspects of the
seismic activity, i.e. the statistics of inter-occurrences or
waiting times of successive earthquakes in an aftershock sequence.
To summarize our results, we have found that these statistics are
consistent with a NHP process driven by a power-law decay rate by
mapping the occurrence of aftershocks in a multi-dimensional space
into a marked point process on the one-dimensional time-line. The
nature of this distribution is closely related to the temporal
correlations between earthquakes and can be used in hazard
assessments of the occurrence of aftershocks. We have also derived
an exact formula describing the distribution of waiting times
between events in a NHP process over a finite time period $T$ and
confirmed it by numerical simulations.

Earthquakes form a hierarchical structure in space and time and can
also be thought of as a branching process where each event can
trigger a sequence of secondary events and so forth. According to
this structure, in some cases it is possible to discriminate between
foreshocks, main shocks, and aftershocks, although, this
classification is not well defined and can be ambiguous. However, it
is observed that moderate and strong earthquakes initiate sequences
of secondary events which decay in time. These sequences are called
aftershocks and their spatial and temporal distributions provide
valuable information about the earthquake generating process
\cite{Kisslinger:1996a}.

Earthquakes follow several empirical scaling laws. Most prominent of
them is Gutenberg-Richter relation \cite{Gutenberg:1954a} which
states that the cumulative number of events greater than magnitude
$m$, $N(\ge m)$, follows an exponential distribution $N(\ge
m)\propto 10^{-b\,m}$, where $b$ is a universal exponent near unity.
This distribution becomes a power-law when the magnitude is replaced
with the seismic moment, as the magnitude scales as a logarithm of
the seismic moment. Another empirical power-law scaling relation
describes the temporal decay of aftershock sequences in terms of the
frequency of occurrence of events per unit time, this is called the
modified Omori's law \cite{Utsu:1995a}. The spatial distribution of
faults on which earthquakes occur also satisfy (multi-)fractal
statistics \cite{Robertson:1995a,Davidsen:2004a}. These laws are
manifestations of the self-similar nature of earthquake processes.

Based on studies of properties of California seismicity, an attempt
to introduce a unified scaling law for the temporal distribution of
earthquakes was proposed \cite{Bak:2002a}. The distribution of
inter-occurrence times between successive earthquakes was obtained
by using as scaling parameter both a grid size over which the region
was subdivided, and a lower magnitude cutoff. Two distinct scaling
regimes were found.  For short times, aftershocks dominate the
scaling properties of the distribution, decaying according to the
modified Omori's law. For long times, an exponential scaling regime
was found that can be associated with the occurrence of main shocks.
To take into account the spatial heterogeneity of seismic activity,
it has been argued that the second regime is not an exponential but
another power-law \cite{Corral:2003a}. An analysis of the change in
behavior between these two regimes based on a nonstationary Poisson
sequence of events was carried out in \cite{Lindman:2005a}. The
further analysis of aftershock sequences in California and Iceland
revealed the existence of another scaling regime for small values of
inter-occurrence times \cite{Davidsen:2004a}.

An alternative approach to describe a unified scaling of earthquake
recurrence times was suggested in
\cite{Corral:2004a,Corral:2004b,Corral:2005a}, where the
distributions computed for different spatial areas and magnitude
ranges were rescaled with the rate of seismic activity for each area
considered. It was argued that the seismic rate fully controls the
observed scaling, and that the shape of the distribution can be
approximated by the generalized gamma function, indicating the
existence of correlations in the recurrence times beyond the
duration of aftershock sequences. This approach agrees with
observations that main shocks show a nonrandom behavior with some
effects of long-range memory \cite{Mega:2003a}.


Before we carry out an analysis of seismic data, we will outline a
derivation of a distribution function for waiting times between
events in a point process characterized by a rate $r(t)$ and
distributed according to NHP statistics over a finite time interval
$T$. The full analysis is going to be reported elsewhere
\cite{Yakovlev:2005a}. The instantaneous probability distribution
function of waiting times $Y$ at time $t$, until the next event in
accordance with the NHP process hypothesis has the following
form~\cite{Daley:2002a}
\begin{equation}\label{nhpm:eq1}
    F(t,\Delta t)=\mathrm{Prob}\{Y < \Delta t\}=1-\mathrm{e}^{-\int_t^{t+\Delta
    t}\,r(u)\,du}\,,
\end{equation}
where $r(t)$ is a rate of occurrence of events at the time $t$. The
probability density function of waiting times over the whole time
period $T$ has the following form
\begin{eqnarray}\label{nhpm:eq2}
    P_T(\Delta t) & = & \frac{1}{N}\,\left[\int\limits_0^{T-\Delta t}\,r(s)\,r(s+\Delta t)\,
    \mathrm{e}^{-\int_s^{s+\Delta t}\,r(u)\,du}\,ds +\right. \nonumber \\
    & & \left.+\,r(\Delta t)\,\mathrm{e}^{-\int_0^{\Delta t}\,r(u)\,du}\right]\,,
\end{eqnarray}
where $N=\int_0^T\,r(u)\,du$ is the total number of events during a
time period $T$. In the simple case of a constant rate
($r=\mathrm{const}$) one recovers the result for the homogeneous
Poisson process, namely, $P_T(\Delta t)= \mathrm{e}^{-r\,\Delta
t}\,\left(r-\frac{r\,\Delta t}{T}+\frac{1}{T}\right)$
\cite{Yakovlev:2005a}.

In order to check the correctness of the derivation of
Eq.~(\ref{nhpm:eq2}) we have performed numerical simulations of the
NHP process with a decaying time dependent rate $r(t)$.
Specifically, we have used a power-law rate defined as
\begin{equation}\label{nhpm:eq3}
    r(t)=\frac{1}{\tau\,\left(1+t/c\right)^p}\,,
\end{equation}
where $\tau$ is a characteristic time that defines the rate at time
$t=0$, $c$ is a second characteristic time that eliminates the
singularity at $t=0$, and $p$ is a power-law exponent. This rate is
commonly used to describe the relaxation of aftershock sequences
after a main shock and is called the modified Omori's
law~\cite{Utsu:1995a}.

\begin{figure}
\includegraphics[scale=0.46,viewport= 0mm 0mm 215mm
120mm,angle=0]{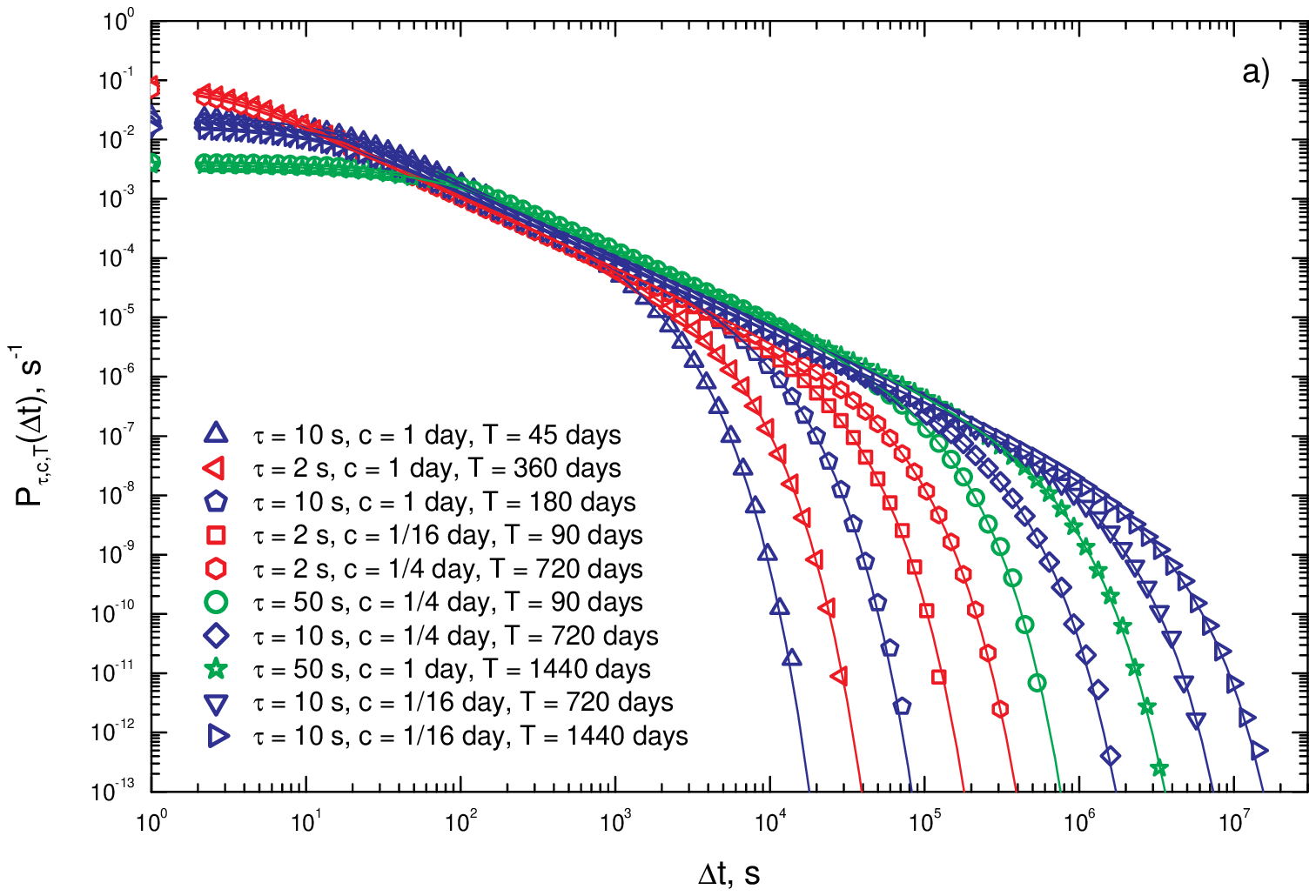}%

\includegraphics[scale=0.46,viewport= 0mm 0mm 215mm
120mm,angle=0]{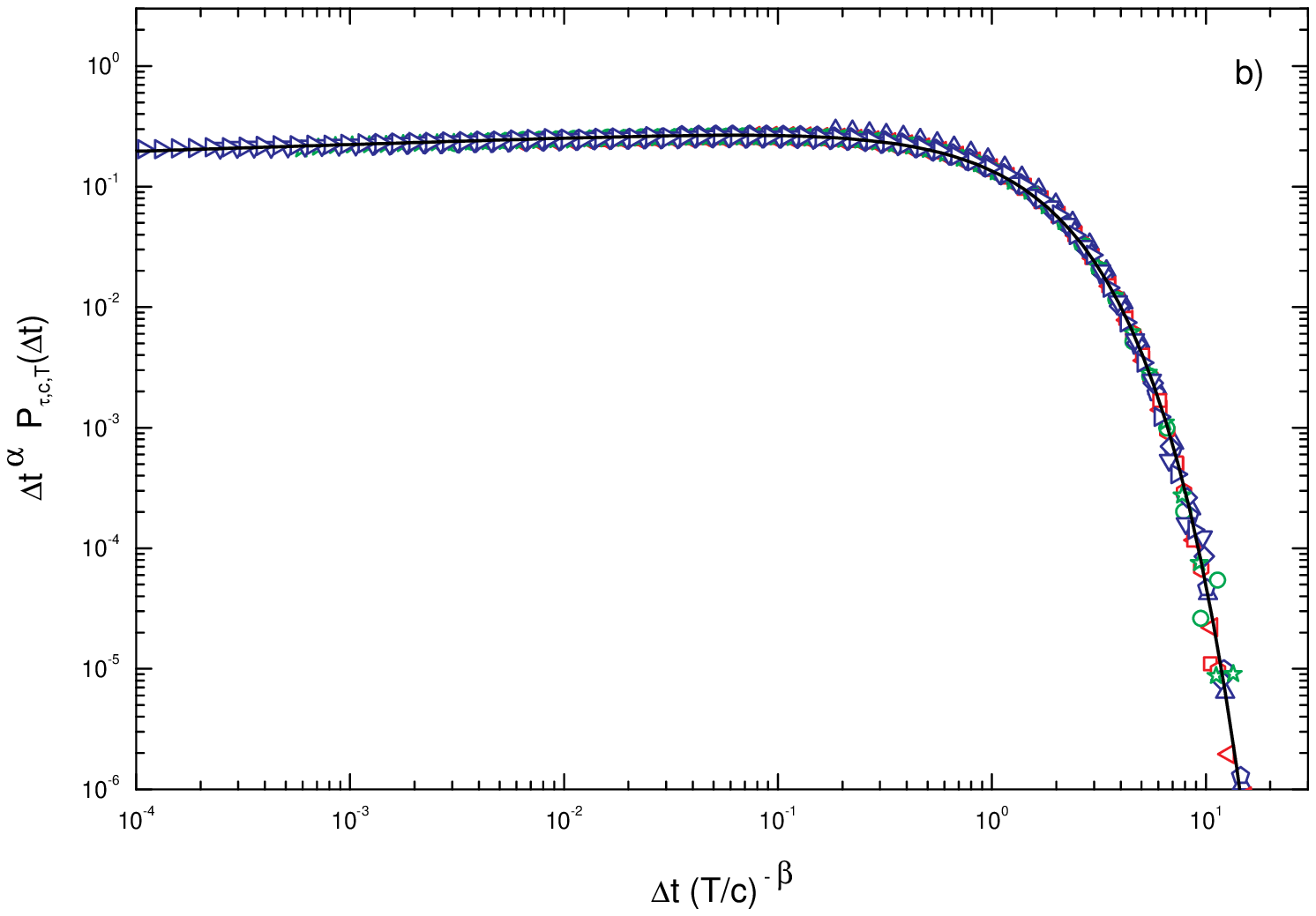}%

\caption{\label{fig1} a) Inter-occurrence time distributions for the
NHP process with the decaying rate (\ref{nhpm:eq3}). Symbols are
numerical simulations of the NHP process. Solid lines are numerical
integrations of Eq.~(\ref{nhpm:eq2}). b) The scaling analysis of the
distributions for $\Delta t\ge 150$~s using Eq.~(\ref{nhpm:eq4}).
The solid line is a fit of the generalized gamma function to the
rescaled data.}
\end{figure}

In Fig.~\ref{fig1}a we show plots of numerical simulations of the
NHP process with varying scaling parameters $\tau$, $c$, and $T$
with fixed $p=1.2$. These are indicated as solid symbols. We also
plot the corresponding numerical integrations of
Eq.~(\ref{nhpm:eq2}) for the same values of $\tau$, $c$, and $T$.
The comparison shows that Eq.~(\ref{nhpm:eq2}) correctly describes
the inter-occurrence time distributions between events.

We have also performed a scaling analysis of our simulated
distributions for the values $\Delta t \ge 150$ s. This is shown in
Fig.~\ref{fig1}b. The distributions collapse onto each other with
respect to the following scaling law
\begin{equation}\label{nhpm:eq4}
    P_{\tau,c,T}(\Delta t) = \frac{1}{\tau}\,
    \left(\frac{\Delta t}{\tau}\right)^{-\alpha}\,
    f\left[\frac{\Delta t}{\tau}\,\left(\frac{T}{c}\right)^{-\beta}\right]\,,
\end{equation}
where $\alpha=1.212$ and $\beta=1.194$. The scaling function $f(x)$
can be approximated by the generalized gamma function
$f(x)=A\,x^{\gamma}\,\exp(-x/B)$ with $\gamma=0.056$, $A=0.329$, and
$B=1.122$. The distribution functions have two characteristic
time-scales, i.e. $\tau$ and $\tau (T/c)^\beta$, which define two
roll overs for small and large $\Delta t$'s.

To find scaling relations between the above exponents, one has to
perform an asymptotic analysis of the integral (\ref{nhpm:eq2}) for
large $\Delta t \gg 1$ and $T \gg 1$. This analysis can be done by
using Laplace's method \cite{Olver:1974a} and gives $P(\Delta t)
\sim {\Delta t}^{-(2-1/p)}$ for $T\to\infty$ and $P(\Delta t) \sim
\exp[-\Delta t/\tau(1+T/c)^p]$ for finite $T$, from which we
conclude that $\alpha-\gamma=2-1/p$ and $\beta=p$. The exponent
$2-1/p$ was also reported in \cite{Utsu:1995a}.

To check the proposed hypothesis that aftershocks can be modeled as
a NHP process, the derived distribution (\ref{nhpm:eq2}) has been
compared to several aftershock sequences in California. We have used
the seismic catalog provided by the Southern California Earthquake
Center (SCSN catalog, http://www.data.scec.org). The identification
of aftershock sequences is usually an ambiguous problem which
requires assumptions on the spatial and temporal clustering of
aftershocks \cite{Kisslinger:1996a}. In this work we assume that all
events that have occurred after a main shock within a given time
interval $T$ and a square area $L\times L$ are aftershocks of a
particular main shock. By neglecting the magnitude, spatial size and
duration of each individual event and considering all aftershocks
above a certain threshold $m_c$, we map a multi-dimensional process
into a process on the one-dimensional time-line. This marked process
is characterized by the times of occurrence of individual events
$t_i$ and their magnitudes $m_i$. We define inter-occurrence or
waiting times between successive aftershocks as $\Delta t_i=t_i -
t_{i-1}$, $i=1,2,\ldots$ and study their statistical distribution
over a finite time interval $T$.

In this work, we use the decay rate of aftershocks, introduced in
\cite{Shcherbakov:2004a,Shcherbakov:2005a}, which is a
generalization of the modified Omori's law (\ref{nhpm:eq3}), where
the characteristic time $c(m)=\tau\,(p-1)\,10^{b\,(m^\star-m)}$ is a
function of the lower cutoff magnitude of an aftershock sequence
$m$, $m^\star$ is the maximum value of an aftershock in a sequence
with finite number of events, and $p>1$. We also assume that
aftershock sequences satisfy the Gutenberg-Richter cumulative
distribution $N(\ge m) = 10^{b\,(m^\star - m)}$. This defines a
truncated distribution where the expected number of events with
magnitudes greater than $m^\star$ is equal to one and the exponent
$b$ is generally near unity \cite{Shcherbakov:2004b}.

\begin{figure}
\includegraphics[scale=0.46,viewport= 0mm 0mm 215mm
120mm,angle=0]{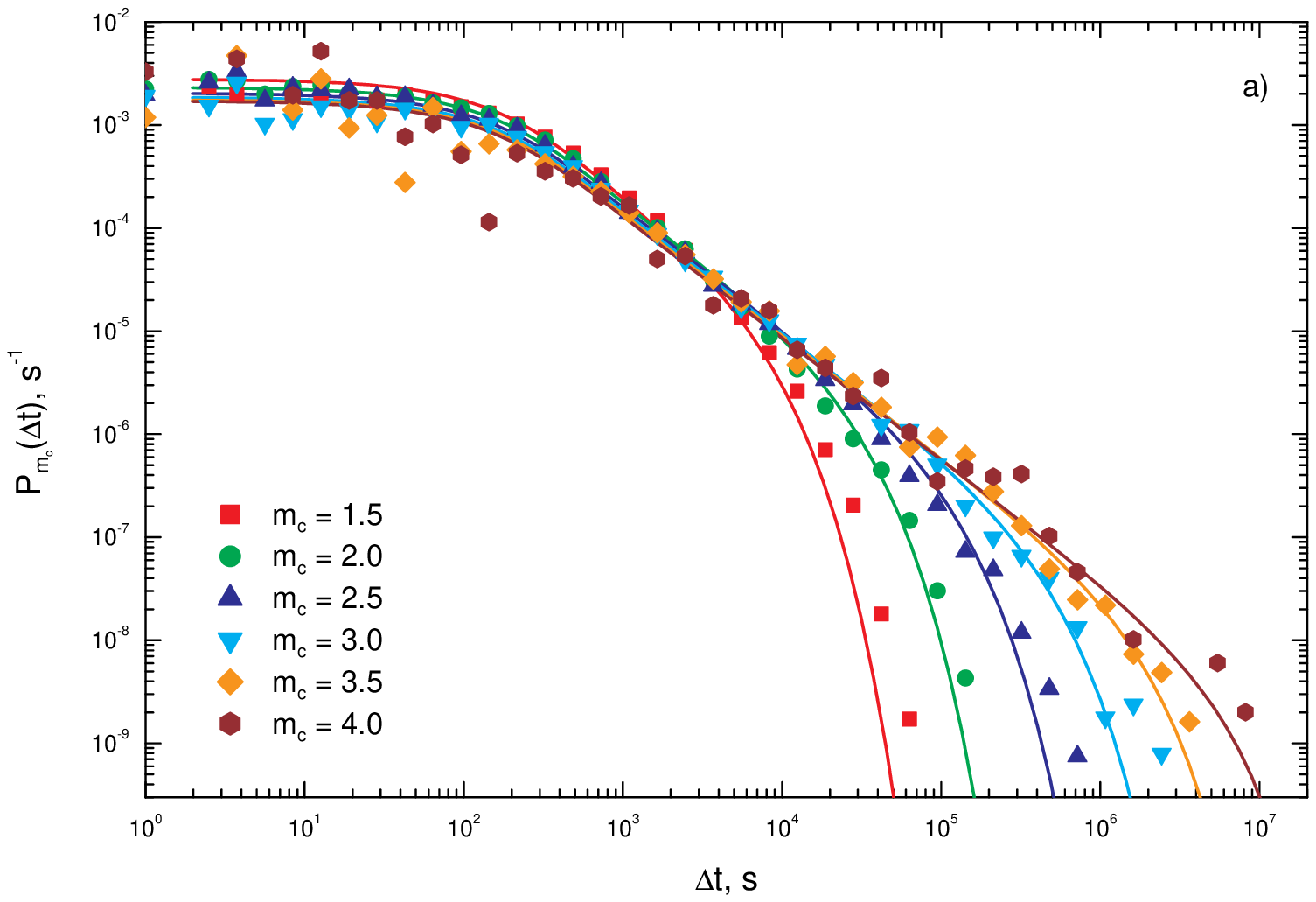}%

\includegraphics[scale=0.46,viewport= 0mm 0mm 215mm
120mm,angle=0]{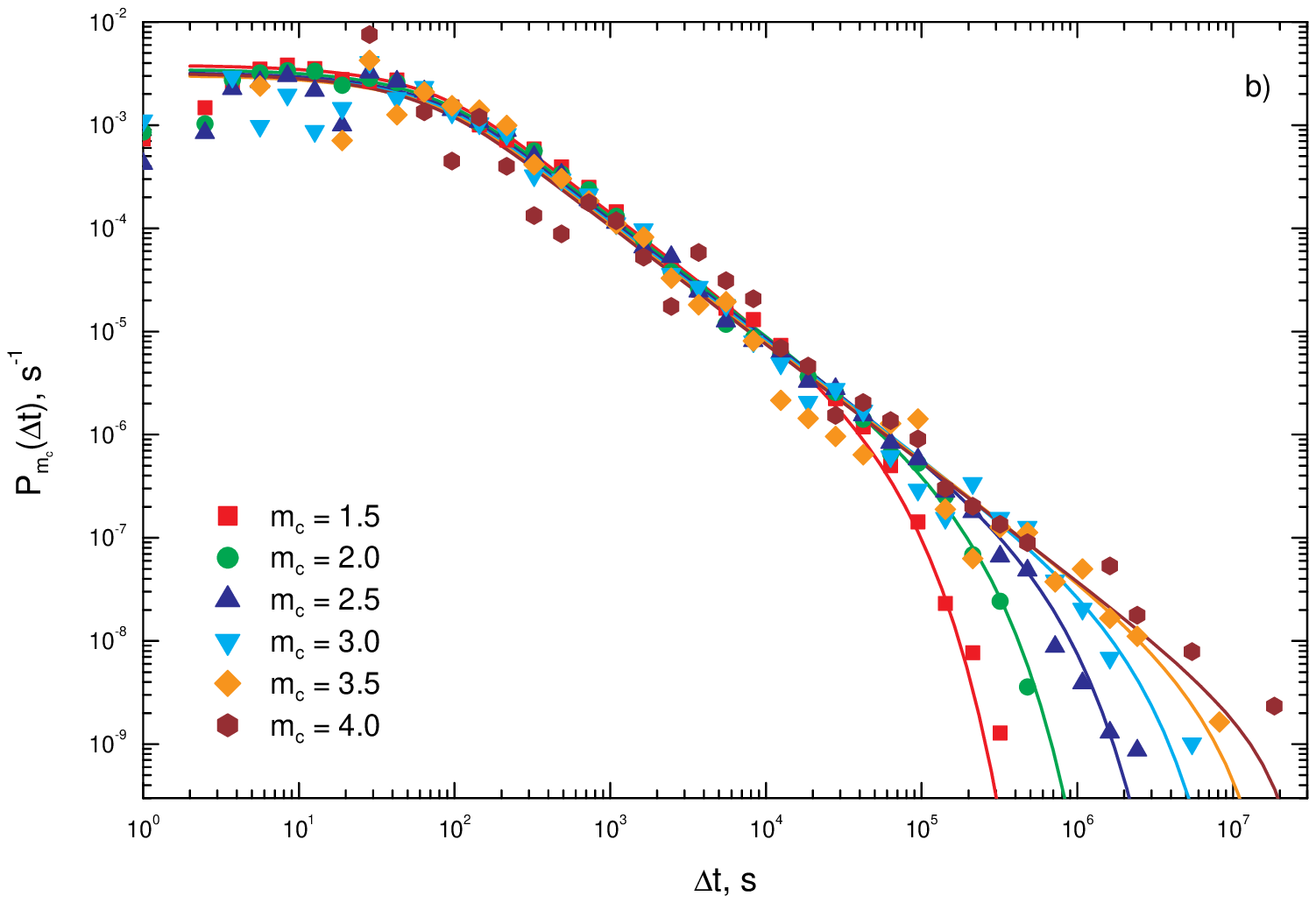}%

\includegraphics[scale=0.46,viewport= 0mm 0mm 215mm
120mm,angle=0]{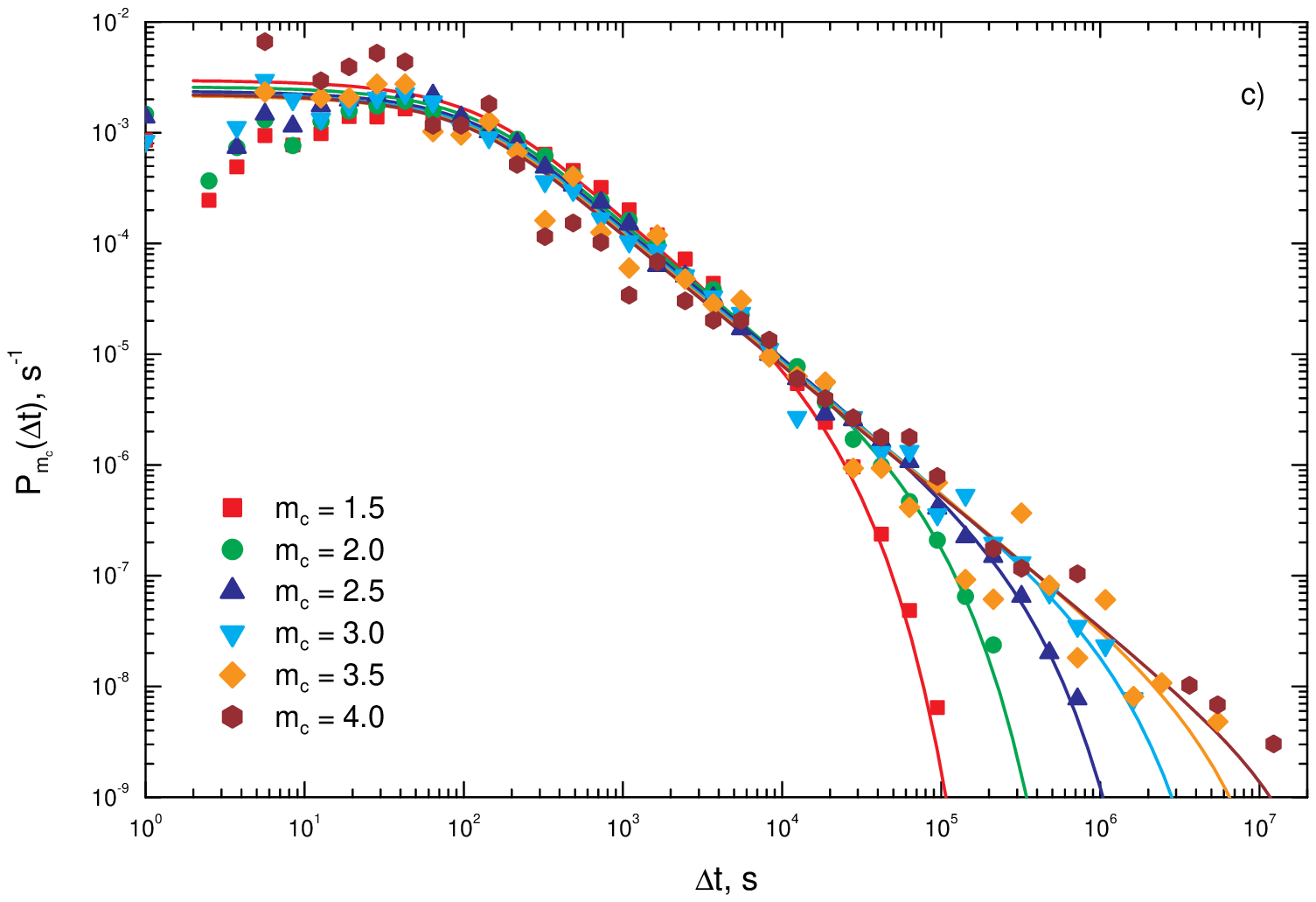}%
\caption{\label{fig2}Inter-occurrence time distributions for the
three California aftershock sequences for different magnitude
cutoffs $m_c=1.5, 2.0, 2.5, \dots, 4.0$. Square areas considered are
$1.25^\circ\times 1.25^\circ$ for (a) the Landers sequence and
$1.0^\circ\times 1.0^\circ$ for (b) the Northridge and (c) Hector
Mine sequences each centered on the epicenters of the main shocks.
In all cases a time interval of $T=1$ year following the main shock
has been used. The solid lines have been computed using
Eq.~(\ref{nhpm:eq2}).}
\end{figure}

In Fig.~\ref{fig2} we have computed the distribution functions of
inter-occurrence times between successive aftershocks from the
observed data of three California aftershock sequences. These are
indicated as solid symbols in Fig.~\ref{fig2}. For each of these
sequences we have used square boxes of size $L\times
L=1.25^\circ\times 1.25^\circ$ for the Landers earthquake
($m_{ms}=7.3$; Jun 28, 1992) and $1.0^\circ\times 1.0^\circ$ for the
Northridge ($m_{ms}=6.7$; Jan 17, 1994) and Hector Mine
($m_{ms}=7.1$; Oct 16, 1999) earthquakes centered on the epicenter
of the main shocks and a time interval of $T=1$ year. All the
earthquakes that occurred in the spatio-temporal boxes were treated
as aftershocks. The analysis of the data shows that the
distributions are not too sensitive to changes in the linear size
$L$ of the box, the results are almost the same for $L$ ranging from
$0.25^\circ$ to $1.75^\circ$ within statistical errors. This means
that the distributions are dominated by the activity of the events
generated by the main shocks and the background seismicity doesn't
contribute significantly to the scaling.

In this analysis of an aftershock sequence as a point process we
treat all earthquakes as having the same magnitude and as a result
we lose a significant fraction of information related to the physics
of the process. To recover some information from the magnitude
domain of each sequence we have used a lower magnitude cutoff $m_c$
as a scaling parameter and study sequences with different $m_c$'s.
These are depicted by different colors in the plots
(Fig.~\ref{fig2}). The distributions with lower magnitude cutoffs
have a shorter power-law scaling regime and start to roll over more
quickly. This can be explained by the presence of more events in the
sequences with lower magnitude cutoffs and as a result the
shortening of the mean time intervals between events. Another
scaling parameter which affects the roll over is the time interval
$T$. An increase in $T$ leads to the occurrence of longer intervals
$\Delta t$ between events.

\begin{figure}
\includegraphics[scale=0.46,viewport= 0mm 0mm 215mm
120mm,angle=0]{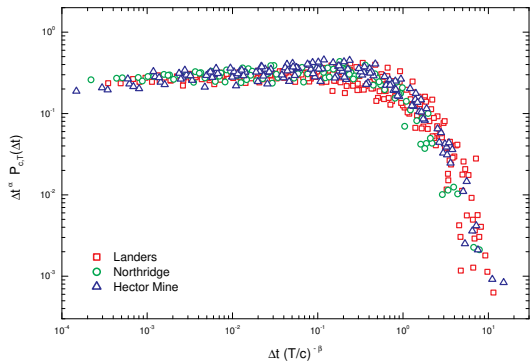}%

\caption{\label{fig3} The scaling analysis of the inter-occurrence
time distributions of the three aftershocks sequences according to
(\ref{nhpm:eq4}). The red, green, and blue symbols correspond to the
Landers ($p=1.22$, $\tau=107.25$ s), Northridge ($p=1.17$,
$\tau=53.14$ s), and Hector Mine ($p=1.22$, $\tau=83.71$ s)
sequences. In each sequence the varying parameters $c$ and $T=180$,
$360$, and $720$ days have been used. The best collapse have been
found for $\alpha=1.25$ and $\beta=1.22$.}
\end{figure}

To compare the observed scaling with the simulations of a NHP
process we also plot in Fig.~\ref{fig2}, as solid curves, the
distributions computed assuming that aftershock sequences follow NHP
statistics with the rate given by Eq.~(\ref{nhpm:eq3}) and the
parameters $\tau$, $c$, and $p$ estimated from the observed three
California aftershock sequences \cite{Shcherbakov:2004a}. The plots
show that the modeled distributions are in excellent agreement with
the observations.

We have also performed a scaling analysis of the inter-occurrence
time distributions of these aftershock sequences. This is shown in
Fig~\ref{fig3}. The characteristic time $c$ and the time interval
$T$ have been chosen as scaling parameters. These sequences are
characterized by slightly different initial rates $\tau$ and
exponents $p$. The results show a reasonably good scaling with
respect to $c$ and $T$ which supports our hypothesis that aftershock
sequences can be described as a NHP process.


In summary, the studies of inter-occurrence of aftershocks presented
in this work suggest that aftershock sequences can be modeled to a
good approximation as a point process governed by NHP statistics,
where the rate of activity decays as a power-law
(Eq.~\ref{nhpm:eq3}). This decaying rate introduces a self-similar
regime into the observed scaling followed by an exponential roll
over. An analysis of a nonstationary earthquake sequence was also
performed in \cite{Corral:2005a}. It was suggested the existence of
a secondary clustering structure within the main sequence and
deviation from NHP behavior. The knowledge of the type of
distribution that governs the occurrence of aftershocks is important
in any hazard assessment programs. The derived distribution
(Eq.~\ref{nhpm:eq2}) has much broader applicability and can be used
for studies of many time dependent processes which follow NHP
statistics.

Fruitful discussions with A. Corral and A. Soshnikov are
acknowledged. The comments of the anonymous reviewer helped to
enhance the analysis. This work has been supported by NSF grant ATM
0327558 (DLT, RS) and US DOE grant DE-FG02-04ER15568 (RS, GY, JBR).

\end{document}